# Darwin Inside the Machines:
# Malware Evolution and the Consequences for Computer Security

*Dimitris Iliopoulos[1], Péter Ször[2], and Christoph Adami[11]*

[1]Keck Graduate Institute of Applied Life Sciences, Claremont, CA 91711
[2]Symantec Corporation, 20330 Stevens Creek Blvd., Cupertino, CA 95014

*Recent advances in anti-malware technologies have steered the security industry away from maintaining vast signature databases and into newer defense technologies such as behavior blocking, application white-listing and others. Most would agree that the reasoning behind this is to keep up with the arms race established between malware writers and the security community almost three decades ago. Still, malware writers have not as yet created new paradigms. Indeed, malicious code development is still largely limited to code pattern changes utilizing polymorphic and metamorphic engines, as well as executable packer and wrapper technologies. Each new malware instance retains the exact same core functionality as its ancestor and only alters the way it looks. What if, instead, malware were able to change its function or behavior autonomously? What if, in the absence of human intervention, computer viruses resembled biological viruses in their ability to adapt to new defense technologies as soon as they came into effect? In this paper, we will provide the theoretical proof behind malware implementation that closely models Darwinian evolution. Biological viruses are under constant attack by immune systems and artificial drugs. Yet they systematically manage to evolve new functionalities that circumvent such countermeasures, leading to recurrent epidemics. According to the biological analogy, evolvable malware will be able to alter its functionality by autonomously incorporating behaviors freely available to it by the numerous discoverable APIs. The new behavior profiles would be constantly screened by security software in the same way natural selection acts on biological organisms. In the end, the malware instances that are better equipped to survive countermeasures will be able to proliferate more efficiently. Such malware pose a real threat to the current methods of detection due to the vast numbers of functions they can adopt and that cannot possibly be screened for. Furthermore, it is likely that clean-program functionality will be favored amongst such behaviors since it shields malware that is mimicking clean programs from behavior blocking. As a consequence, we predict that behavior-based virus detection would quickly become ineffective if malware can evolve based on the Darwinian paradigm.*

During 2007 alone, attackers created over 700,000 new malicious programs [1,2]. While this number appears daunting, modern antivirus (AV) systems have dealt with this crop successfully every year, by using sophisticated signature and behavior detection techniques. All of the new viruses are, as far as we know, the product of human engineering. Here we discuss the possibility of an alternative form of design: the emergence of novel function in an autonomously evolving piece of malware.

The possibility of autonomously evolving computer programs is not new. Artificial Life ("Alife") researchers have studied self-replicators in computer environments since 1979 [3], and the first autonomously mutating self-replicating computer programs were introduced by Ray in 1992 [4]. More recent systems such as the Avida platform have established that Darwinian evolution of computer programs can be used to study evolutionary biology and genetics [5,6]. As opposed to computer malware, Alife systems are squarely aimed at the research environment. In these systems, the code is implemented with small, well-defined instruction sets that are highly evolvable, that is, the probability that a mutation leads to another functional program is high (of the order of several to tens of percents). Most standard computer languages are not robust in this manner. For example, the x86 instruction set only tolerates very few mutations (on the order of fractions of a percent of the code [7]). Alife systems also differ in that they implement *simulated* environments where specific program behaviors are rewarded. While this setting is very different from the standard computer malware paradigm, computer virus researchers hinted at the idea of autonomously evolving malware already in the early 1990s [8]. In particular, Spafford performed an exhaustive review of computer viruses under the Artificial Life perspective confirming the absence of functional evolution in such programs [9]. Although the possibility of an autonomously evolving virus—as the one discussed here—was mentioned, it was quickly dismissed as a task too daunting requiring a very large implementation, possibly larger than the OS itself. More recently, the idea has resurfaced, with a higher emphasis this time on the outcome [10] and the possible mechanics of its emergence [11]. We argue here that apart from a hint and some failed implementation attempts (e.g., W32/Zellome, see [12]), self-evolving malware has yet to appear. The concept itself, however, is relatively simple, and the consequences of the release of evolving malware should be studied.

Darwinian evolution versus malware evolution

Evolution as a process requires the presence of three simple elements: *replication*, *variation* and *differential fitness*. Replication allows for inheritance, and in particular the inheritance of variations, introduced by different mechanisms of mutation. These variations are then selected for or against via the competition of individuals over limited resources. The competition leads to differential fitness, i.e., differences in reproductive success between individuals within a population. Mutations that provide organisms with a reproductive advantage over others (beneficial mutations) will tend to propagate to future generations. It is important to note that the fraction of beneficial mutations (among all those that can occur) is usually very small, with most mutations being deleterious and the rest neutral. Nonetheless, given the long time scale available to biological evolution along with large population sizes, such gradual stepwise beneficial changes will lead to the emergence of complexity [13]. Some have argued that neither the geological timescales nor the large populations that are germane to biological evolution will be available to evolving computer malware. However, generation times for computer viruses are many orders of magnitude shorter than even bacterial generation times, and furthermore selective pressures are expected to be much stronger and mutation rates higher, speeding up evolution [14]. We should note here that geological times, in contrast to popular belief, are not a prerequisite for the appearance of evolutionary change, even in biological evolution. The emergence of new functional complexity on much smaller time scales than geological ones has recently been observed in very different experiments. A 36-year experiment with lizards for example has shown that such a comparatively short time frame is sufficient for major adaptive morphological changes to arise, such as novel morphological structures in the form of cecal valves [15]. Also, in a long-term experiment with *E. coli* bacteria, one particular strain evolved the ability to metabolize a new carbon source (citrate) after about 31,500 generations, a complex adaptation that required a multitude of mutations [16]. Given the drastically shorter generation time of computer viruses, generations of this order of magnitude can be achieved within weeks at most.

During the three decades of their existence, computer viruses have moved from simple replicators to advanced polymorphic and metamorphic implementations [17]. The underlying goal of this progression has been to increase the variability of the virus' signature to the point that tracking different instances of the same virus becomes too daunting a task. Nonetheless, even for the most sophisticated metamorphic viruses [18], the specific functionality and overall behavior of the virus remain intact. Signature obfuscation, or as we will refer to it here as

"cryptic variation", will not allow for the discovery of new functionality. What true Darwinian evolution can accomplish is vastly different, because it is the process responsible for the *de novo* generation of all of the complexity of life.

The variation observed in surviving lineages of biological viruses (as compared to their ancestors) is a direct result of information "exchange" between the virus and its environment. Simply put, biological viruses are constantly testing new ways of exploiting environmental resources via the process of mutation. In contrast, computer viruses do not exhibit such traits, relying instead on changing their appearance to avoid detection. *Functional* (as opposed to cryptic) variation, such as the discovery of a new exploit or the mimicry of non-malicious behavior masking malicious actions, is not part of the arsenal of current malware. While there are examples of functional variation that have occurred by chance (reviewed below in support of our hypothesis) there are no examples of computer malware that exhibits intentional functional change between generations. In the absence of functional variation, differential fitness will never be realized in computer viruses since the reproductive success of offspring remains unchanged. In the event that a behavioral signature is developed for the virus, the entire population, including the cryptic variants, is affected equally. In this case, the functional/behavioral uniformity of the virus population would force all of it to extinction by the countermeasure. Functional variants, on the other hand, because of the variation in behavior, can escape behavior-based detection methods. Below, we investigate the theoretical possibility of functional variation in computer viruses and their consequences.

## Model implementation

Suppose, malware M is comprised of an arbitrary number of malicious functions:

$$M = \{M_1, M_2, M_3, ..., M_N\} .$$

An evolutionary function $EF \in M$, is introduced into the existing set:

$$M = \{M_1, M_2, M_3, ..., M_N, EF\} .$$

EF has the ability to generate new functional code N after *each* generation of M, with a given probability. This can be achieved by different methods, including via the insertion of code extracted from randomly selected APIs present in the malware's native environment (for example, extraction of function calls that are part of the Windows API). Alternatively, in the case of a script threat, extraction of script code from other scripts might also serve the same purpose. In any case, the resulting malware M now consists of the set

$$M(EF) = \{M_1, M_2, M_3, ... M_N, EF, N_1..N_m\}, \text{ where } N=N(EF).$$

The newly generated function N need not necessarily be malicious. Also, the $M_i$ might trigger antivirus responses individually or as a combination of each other. The malware M(EF) is now an evolvable threat.

Researchers that envisioned how autonomously evolving malware could be coded have largely focused on binary code manipulation [8,11]. The high percentage of lethal mutations that would be experienced by such malware (due to the brittleness of the code) would forbid evolution from occurring. Consequently, the need for an "evolvable" language or meta-language of implementation becomes apparent. While some work in this direction has been undertaken [7], the prospects for such a design are daunting. By relying instead on functional code that is already present in the malware's environment, the problem is reduced to finding a way of adopting that functionality in the malware code, instead of creating a language that can code for that functionality autonomously.

Another major consideration with such an implementation is the actual coding of EF itself, as it might appear to represent a static component for which a signature could be produced. However, the EF could use any number of advanced polymorphic or metamorphic engines to hide itself from detection. The ongoing effort by the antivirus community to avoid maintaining the large signature databases of the past, combined with the advanced metamorphic engines used by viruses such as W95/Zmist [19] make a cryptic EF implementation feasible. In fact, it is highly probable that EF's presence can be fully disguised by M. By simulating non-malicious behavior for example, malware would at best be classified as a false positive even if EF is detectable. Given the antivirus industry's aversion to false positives, it is conceivable that malware would exploit this loophole and autonomously and forcefully mimic clean application behavior.

## Functional evolution in current security environments

Adapting to security environments can be illustrated via a number of examples. In this section, we list several ways in which random changes can lead to evasion of detection from behavior- and signature-based blocking as well as application white-listing.

**Clean-application mimicry.** AV systems are usually designed with a bias to avoid false positives. This bias creates a loophole that can be exploited by functional adaptation, because as the number of detections by an AV system increases so do the false positives of each product. As a consequence, the AV industry attempts to reduce the footprint of signatures, but this can significantly amplify the risks of exploitation by programs that have learned to mimic clean applications either by behavior or by signature.

Imagine that a program imports API function "Foo", which makes it similar to programs that create false positives for the AV system. As a result, the presence of function Foo excludes that program from the attention of behavioral blocking. This problem exists even if the system uses weighting, as long as the negative weights to identify possible clean applications can be simulated by the newly evolved functions in malware M. There are analogies for this type of evolutionary adaptation in biology. Some viruses have evolved a set of proteins that mimic precisely those proteins that are involved in combating the virus (the so-called complement regulator proteins [20]), or else by mimicking a protein that the immune system uses to recognize viruses (so-called Fc proteins [21]). In both cases, the immune system is fooled into treating the virus as "clean" because it looks like part of the immune system itself.

If malware M appears to have an application user interface, it will most likely be classified as a clean application because typical malware does not have one. In another example, if malware M is *not* packaged or wrapped, it might appear much less suspicious. Malware writers use packers extensively to hide signatures from AV scanners. One of the cheapest ways for attackers to do this is to use run-time packers or wrappers on top of the existing malware. As a result, malware released in packed form might look much more suspect than a non-packed version. Evolving malware capable to present itself in both packed and unpacked forms would create new challenges for heuristics. There is a parallel to this evasion technique in biological viruses. In fact, often a body's immune system triggers on the envelope (or capsid) of a virus. As a consequence, some viruses have adapted to an infection mode without an envelope and capsid (for example, the viroids [22]). The reverse strategy also exists, where viruses such as the herpes viruses, in an attempt to masquerade themselves as host cells, create elaborate envelopes from host cell membranes that contain a series of host-unique markers on them. Such envelopes allow the herpes virus to evade host detection and suppress immune responses

[22].

**White-listing deception.** The practice of application white-listing offers another exploit for evolvable malware. Because the number of known applications is already very high, typically application white-listing only focuses on executable applications (such as Portable Executable files on Windows systems). As a consequence, attacks originating from other types of objects (such as documents) are not controllable by these systems. Malware that is not presented in an executable form would thus be excluded from the attention of white-listing systems completely. Self-evolving executable malware that could convert itself to new forms, for example via executable-to-script-conversion, would pose a challenge to any white-listing system. Instead of presenting itself as an executable, the code might rewrite itself as a macro in a document, or as a command-line script.

Another problem appears if there are no file objects involved, as in the case of the W32/CodeRed family that target a white-listed application in memory over the network. Not surprisingly, when such threats emerge for the first time, they often cause an epidemic. While in-memory threats are not common, their presence is expected to rise if there is a need to adapt to a known white-listing solution.

Another scenario involving white-listing deception could be the following. Suppose a program creates large populations of clean programs that it distributes over the web. However, the clean programs are designed in such a way that their MD5 hash is identical to its own (malicious) self [23]. Because white-listing applications collect executables from a variety of "trusted" sources, the wide distribution of clean "twin" programs could have the result of white-listing the malicious version. Thus, such an attack can exploit both application white-listing as well as application reputation systems, which score applications based on their popularity among users with good reputation. A simple biological example of such an attack is represented by the evasion of the DNA methylation defense of bacteria. Bacteria tag their own DNA with a methyl group, and cut up any DNA inside the cell that does not carry such a group (the methylation is the analogue of an MD5 hash). Viruses have emerged that have learned to emulate these tags (white-listing themselves in the process) and thus escape detection.

**Virtual Machine anti-emulation.** It is conceivable that the size of a long function loop, or a function that calculates parameters to function calls on the fly, could be artificially increased by malware. In such a case, the resulting novel functions N might accidentally create an anti-emulation feature for a virtual machine by exhausting the emulators preset upper limit iteration number to examine the program. This way, M will be capable to compete with an emulator, thus exploiting this feature for its own survival. In this case, the threat will be capable to circumvent not only behavioral blockers that rely on emulators, but also a scanner's built-in heuristics analyzers as well. Random sequences of APIs with parameter-check can also introduce such issues. This trick is often used by existing polymorphic threats as an anti-emulation feature. The W95/Drill family of viruses utilized random calls to a predefined set of APIs to confuse emulators. Later on, the Storm Worm attacks (that created the Storm botnet) delivered thousands of copies of different malware executables that utilized layers such as W32/Tibs on them to confuse antivirus products. The API sets used were changed by the attackers regularly to reduce chances of detections based on the presence of such an API profile. A self-evolving threat would carry this feature by its own nature, as it can create new code, and verification functions for them. We are not aware of a biological analogy for this trick.

**Evasion by proxy.** Even security products such as personal firewalls can be affected by self-evolving threats. Suppose threat M has a feature to proxy behavior using another executable. While the threat may have several malicious features, it has the option to either execute them itself, or by using a running application as a proxy instead. As a result, when executing the communication via a trusted application, the threat can remain undetected when communicating to an outside location (such as the command control channel of a bot network). Alternatively, all malicious features might be separated among several processes as threads. Imagine that a set of threads from a set of processes together constitutes a malicious program. Yet, on its own, each feature might not be triggering the attention of heuristics or the behavior blocking engine, allowing the threat to escape attention.

Trust relationships can also introduce problems. For example, Windows Vista does not allow an executable to alter certain file permissions, even if an administrator executes the application. Still, using a script such permissions can be changed. For example, files such as kernel32.dll and their permissions can only be changed by so called *Trusted Installers* who have full control over these files (see Appendix A). If one tries to change the permission using CACLS.EXE from the Resource Kit, the attempt will fail because this executable cannot be used to take ownership of files, since it is not being trusted. Yet, one can use XCACLS.VBS, and execute it with CSCRIPT.EXE. The script is trusted, since the script interpreter is a trusted application. This means that threats can potentially evolve by changing their representative forms (such as the particular language environment they use) to develop adaptive features to a security environment. What an executable is not able to do due to security restrictions might be achievable as soon as the threat feature is executed as a script. In addition, using a trusted proxy such as EXPLORER.EXE on Vista, a threat can take ownership of files that otherwise it would not be allowed to do (for example using code injection techniques to become part of the process address space of Explorer in memory.)

**Environment-induced functional variation.** New functionality can emerge as a result of environmental effects. Many web sites, for example, change the code within HTML files on the client side arbitrarily by presenting extra functionality to the user. In particular, links to advertisement messages can be replaced with actual content. In other cases, scripts are inserted into pages, and changed on an ongoing basis (see Appendix B.) Such effects could change actual malware code, rendering AV scanners incapable of identifying the threat.

We have seen several verified cases in the past where random corruptions to virus code resulted in a new variant of the virus that renders it undetected by security software. For example, a variation of the W95/CIH family exists that is the result of such a corruption: the replica escaped the attention of most scanners when the natural modification happened to a piece of its nonessential code, without altering the threat's replication functionality. In another case, an error in Microsoft Word resulted in the creation of thousands of macro virus variants via random corruptions. These macro-body corruptions were often ignored by the macro interpreter due to the existence of error handlers in the original virus code. As a consequence, these Word-induced variations could easily result in surviving mutations, as long as a single MacroCopy() command responsible for the successful replication of the threat was still present unchanged.

**Malware code merging.** A threat might be able to snatch code from another program in its environment. We have seen examples of a virus like *Pinfi* jumping on top of worms to replicate in new environments as a combination threat. Security products do not always recognize the worm once it is infected with a virus, and the combination helps the survival of both threats. Disinfected worm copies can also escape attention, as can copies of worm replicas that changed due to transfer of code over network channels. It is conceivable that an evolutionary function in the malware could snatch clean or malicious code from other programs. It could integrate code from other programs by identifying function prolog and epilog code. When this takes place, a function is safely inserted into the code base of the evolutionary virus as a new "function" by running the newly acquired code as a new thread. Existing features might be replaced by the code, which could end up producing reliable output to a given input. (For example, as long as function X returns values greater than 0, it is accepted.) Even complete functionality might be snatched from another clean program, or another virus as well. As previously predicted [17], a cooperation protocol can enhance sharing of features between malicious executables as well. Code snatching is a tried and true function of almost all biological organisms. Bacteria exchange code in small segments called *plasmids*, while viruses routinely integrate bacterial code into their own. Often, viruses carry this piece of code to other bacteria, a phenomenon known as *transduction*.

A form of evolution was observed in macro viruses, which often merge their code base into a document. (Note that the integration of viral code into the host code is the default action for several biological viruses, e.g. HIV [24].) Often the file has a clean macro, and a virus with a set of macros. In addition, another virus may insert its set of macros at the same time, leading to viral macro code merging with both viral and clean macro code. In biology, this phenomenon is quite common, and known as *coinfection*. For example, during a mode of viral reproduction termed "lysogenic", viruses such as bacteriophages integrate their genome into the host's genome and become dormant. When such viruses convert back to the *lytic* (aggressive) mode of reproduction, they excise themselves from the genome. In the event of coinfection, the excised viral genome might be a combination of more than one viral strain leading to the creation of chimeric viral particles. Taking these lessons back to computer malware, we predict that while this merging of code is still manageable for the antivirus programs, cases when a clean macro is merged with a parasitic macro virus might not be. Imagine a random change that results in the macro code becoming very large, either as the result of several layers of merged code or additional polymorphisms. This can easily exhaust typical scanning engines for macro viruses due to memory limitations reserved for the macro code within the engines. As a consequence, an evolved copy can potentially escape just by being too large.

Code snatching can also lead to detection evasion if the snatched code is clean, and triggers the heuristic based on code snippet based exclusions, leading to effects similar to those mentioned under the heading "Clean application mimicry". Once a copy of the malware snatches the code snippet "exclude" from a clean program and presents it within itself at the right place, the scanner might ignore the malware. While the scanner would trigger on the rest of the code, the exclusion forces it to fail. Such exclusion strings were extracted from several AV systems in the past, and shared among attackers to include it in viruses [25]. As this example shows, it is not necessary for a human to disassemble the antivirus programs to gain this knowledge: random code snatching can easily produce it. And once it is produced it will flourish, as it will escape detection.

## How far can evolution go?

We have reviewed numerous examples of non-evolution-induced adaptations to security environments, leading to unintended functional gains by malware. The question that arises from these examples is whether small malware code changes—even if collected under a single malware implementation—are sufficient for the evolution of the functional complexity exhibited by biological viruses. Let us first consider how much functionality and complexity can possibly be generated by the process of random changes and subsequent selection. Biologists often use the vertebrate eye as an example of how complexity can arise through evolution. It is rather difficult to imagine such an organ with its hundreds of millions of photocells and hundreds of different chemicals required to process a single image being created from scratch. Instead, it is easier to imagine a historical succession of eyes that differ only slightly from each other, and where one originates from the other via only small gradual changes or mutations. In a collection of all possible mutations a version of the eye can undergo, large ones are statistically more improbable than smaller ones. Hence, a long string of such small mutations occurring gradually through geological time could connect ancestral versions of the eye to more recent highly functional ones that have the look and feel of design. If we were to follow all these changes through time, it would become apparent instead that ancestral versions are only slightly different from their direct descendants. Indeed, we can see just such a succession of changes when looking at the vast variety of eyes and eye-like structures present in surviving species. From pigment enriched light-sensitive spots in some single-celled animals to elementary eye-like pigmented cavities in several shellfish, the evolution of eye structures can be mapped out [26]. By analogy, we can imagine the emergence of arbitrary complexity in autonomously evolving computer viruses. The myriad examples mentioned above of unintended virus behaviors due to random changes are all examples of more complex behaviors originating from simpler ones through small programming jumps. As in biology, the collection of software and APIs in an operating system environment is sufficient for providing a pool of features that a virus can choose to sample. Those features that provide the virus with an evolutionary advantage (e.g., escaping detection, infecting different platforms, etc.) will be propagated, leading to unanticipated complex behaviors. The only question that remains is whether such a sequence of events is likely within time frames much shorter than the geological time available to the biological process. Lenski et al. have shown that complex functionality can evolve from much simpler functions in Artificial Life platforms, over the course of a few days [13], and complex adaptations have been seen to emerge after tens to tens of thousands of generations [15,16], easily within range of weeks of malware evolution.

Research in Artificial Life of the sort mentioned earlier has identified a number of important parameters that need to be satisfied before complexity can evolve. First and foremost are of course the three conditions for Darwinian evolution themselves: replication, variation and selection. Of those, replication and selection are naturally present, and we argued above that variation can occur via a variety of mechanisms. However, in order for sufficient variation to be present so that selection can act efficiently, population sizes should be large. Finally, the number of variants (mutants) with non-zero fitness (sometimes called the neutral fraction of mutants) should be sizable. It has been shown that ideally between 10%-40% of all mutations experienced by a population need to be neutral to ensure evolvability [7]. (Note that the fraction of mutants of biological proteins is also in that range [27].) Neutrality enables the organism to better sample the pool of available mutations by being able to discover beneficial mutants without constantly being penalized by the already high frequency of deleterious mutations. Beneficial mutations are hard to come by in a single try, but can become more likely if interspersed by a series of neutral ones. In essence, neutral mutations allow organisms to buy more time in their quest for discovering survivability-enhancing traits. Applying this lesson back to computer language implementations, it is easy to see why mutating binary code would make it rather impossible to discover new functionality: the fraction of neutral changes is just too low. On the contrary, integration of pieces of functional code and/or importing functions and APIs available in the computer environment have a much higher probability of being neutral, and very rarely even beneficial.

Through several experiments it was shown that Alife organisms could indeed evolve complexity as long as the intermediate steps that lead to it could be sampled, i.e., the simple functions that are required to build more complex ones can be attained via mutations [13]. It is also interesting to note that this complexity was attained by different populations in a variety of environments, hinting at what biologists refer to as convergent evolution. If a trait produces a significant survival advantage to an organism (be it eye, wings, lungs etc.), it *will* be evolved by unrelated organisms that occupy different environments, in one form or another. What does this mean for computer malware? If computer malware are implemented using the principles of Darwinian evolution and an implementation construct allowing sufficient neutrality, the emergence of complex adaptive behaviors becomes an expected result rather than an improbability, as long as exploitable opportunities exist within the malware's environment. We have here listed a number of those, and are certain that many more can be imagined.

## Conclusion

It has already been discussed by other researchers that no virus detection algorithm can detect all possible viruses, known and unknown [28,29]. The same researchers have also mentioned the possibility of a Darwinian evolution-based malware implementation. So far, no one has attempted to investigate how feasible such an implementation is or assessed the implications it could have, given the current methods of detection. We have shown here through a series of historical malware examples as well as a theoretical implementation model that a truly undetectable virus might be more feasible than previously imagined. By using an evolutionary function, computer malware could implement traits and tricks that allow them to mimic clean application behavior, have themselves white-listed, avoid signature-based detection, or recruit proxies to do their bidding. More importantly, malware could conceivably alter its functionality so as to discover new exploits or find new platforms to infect, with the same ease that biological viruses gain immunity to antiviral drugs, or adapt to new hosts.

The development of sophisticated encryption techniques and metamorphic viruses has led to weakened performance of AV systems based on databases of signatures alone. Behavior-based detection, (relying instead on libraries of behaviors obtained from emulation or from runtime code on the system) have been developed as a stop-gap measure. If malware is developed that can evolve via Darwinian principles, behavior-based blocking is weakened too, as the discovery of even trivial new functionality would have the potential to evade detection. Would there be no defense against such viruses? The answer to this question is currently unclear. The general problem of computer virus detection is inherently an *integrity* problem, and thus managing integrity is essential for the protection of the system. Yet, as the examples for evolution of malware threats have demonstrated, over time the definition of integrity may need to be continuously updated. On the one hand, it is clear that the emergence of an evolvable threat could be potentially disastrous. On the other hand, it is entirely possible that evolution will in most cases lead to harmless forms of a threat, as is often observed in biology. Finally, even though both the signature and the behavior of an evolvable threat may be highly variable, it is still possible that such threats rely on common algorithmic features for which detection methods can be developed. What these commonalities are is, of course, presently unknown.

## Acknowledgements

We would like to thank N. Chaumont, P. Ferrie, V. Griffith, A. Hintze, A. Nash, C. Ofria, and M. Rupp, for discussions and comments on the manuscript. This research was supported in part by the Seaver Foundation.

## Appendix A - Permissions of kernel32.dll on Vista

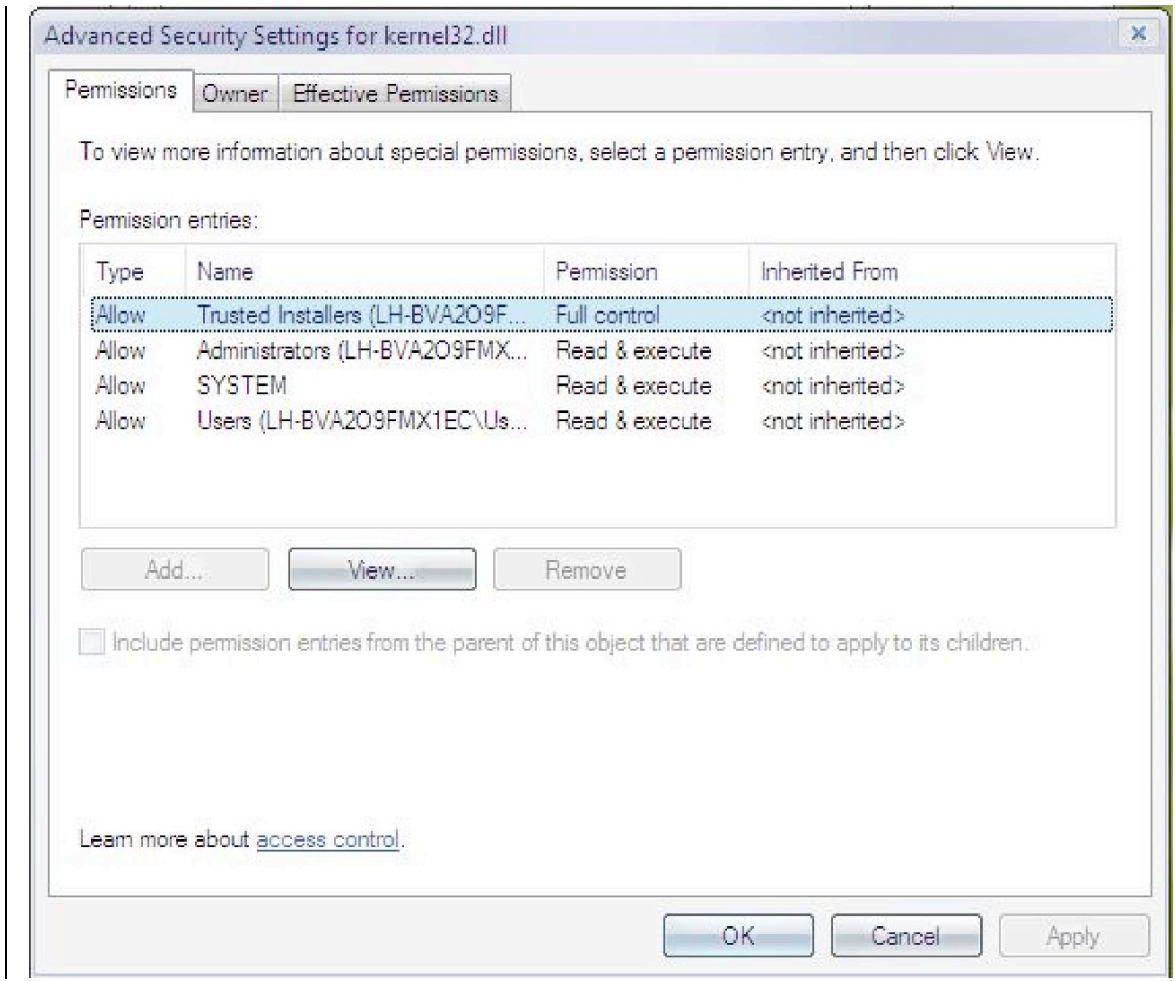

## Appendix B

a.) Viewing the main page of Yahoo at 11:55:15

```
    <script
language=javascript>ULT_KEY='X3oDMTUxcGV1M3FzBGNjA3VzBHlwdWxzZQNjb25zZXJyBHB
hAy0xBHBjaWQDTmV3c19Nb3N0X1JlY29tbWVuZGVkX2VsaV9tYW5uaW5nX3N1cGVyYm93bF81Qi0
3MDc5NDEEcHBpZAMxMjAyMTEwMjgwBGZwY2sDUjZkdHFnRUEEdG1wbANpbmRleC1sBF9TAzI3MTY
xNDkEcGlkAzEyMDIxNTQzNjMEdGVzdAMw';</script><!-- pbt 1202154363 --><script
language=javascript>

    if(window.yzq_p==null)document.write("<scr"+"ipt
language=javascript
src=http://l.yimg.com/us.js.yimg.com/lib/bc/bc_2.0.4.js></scr"+"ipt>")
;
```

```
        </script><script language=javascript>

    if(window.yzq_p)yzq_p('P=Cyq4LtGDJJVLM60kR6dtL9SZDJs6tUenbaMAA..D&T=13tld
je64%2fX%3d1202154915%2fE%3d2716149%2fR%3dyahoo_top%2fK%3d5%2fV%3d1.1%2fW%3d
J%2fY%3dYAHOO%2fF%3d1101005185%2fS%3d1%2fJ%3d952483D1');

    if(window.yzq_s)yzq_s();

    </script><noscript><img width=1 height=1 alt="" src="http://
us.bc.yahoo.com/b?P=Cyq4LtGDJJVLM60kR6dtL9SZDJs6tUenbaMAA..D&T=1420ruvld%2f
X%3d1202154915%2fE%3d2716149%2fR%3dyahoo_top%2fK%3d5%2fV%3d3.1%2fW%3dJ%2fY
%3dYAHOO%2fF%3d1142761950%2fQ%3d-1%2fS%3d1%2fJ%3d952483D1"></noscript>

    <!-- f22.www.sp1.yahoo.com compressed/chunked Mon Feb 4 11:55:15 PST
2008
-->
```

b.) Viewing the main page of Yahoo at 11:56:17. Differences to previous page are in bold.

```
    <script
language=javascript>ULT_KEY='X3oDMTU1M2I4YzljBGNjA3VzBHlwdWxzZQNjb25zZXJyBHB
hAy0xBHBjaWQDT01HX1BvcHVsYXJfQ2FuZGlkX0NlbGViaX1Bob3Rvc19qZXNzaWNhX2FsYmFfVE5
fM0UtNzA3MDE0BHBwaWQDMTIwMTg5NDA0OARmcGNrA1I2ZHRxZ0VBBHRtGwDaW5kZXgtbARfUwM
yNzE2MTQ5BHBpZAMxMjAyMTU0MzYzBHRlc3QDMA--';</script><!-- pbt 1202154363 -
-><script language=javascript>

    if(window.yzq_p==null)document.write("<scr"+"ipt
language=javascript
src=http://l.yimg.com/us.js.yimg.com/lib/bc/bc_2.0.4.js></scr"+"ipt>")
;

    </script><script language=javascript>

    if(window.yzq_p)yzq_p('P=HLySgUS0zpCOOnAXnkEEPL8PDJs6tUenbeEABiC8&T=13t70
v096%2fX%3d1202154977%2fE%3d2716149%2fR%3dyahoo_top%2fK%3d5%2fV%3d1.1%2fW%3d
J%2fY%3dYAHOO%2fF%3d3065009760%2fS%3d1%2fJ%3d90CEB444');

    if(window.yzq_s)yzq_s();

</script><noscript><img width=1 height=1 alt=""src="http://
us.bc.yahoo.com/b?P=HLySgUS0zpCOOnAXnkEEPL8PDJs6tUenbeEABiC8&T=140i4ljqj%2fX%
3d1202154977%2fE%3d2716149%2fR%3dyahoo_top%2fK%3d5%2fV%3d3.1%2fW%3dJ%2fY%3dYA
HOO%2fF%3d4228518O%2fQ%3d1%2fS%3d1%2fJ%3d90CEB444"></noscript>
    <!-- f41.www.sp1.yahoo.com compressed/chunked Mon Feb 4 11:56:17 PST 2008
-->
```